\documentclass[prb,twocolumn,showpacs]{revtex4}
\usepackage{graphicx}
\usepackage{dcolumn}
\usepackage{bm}
\usepackage{amsmath}
\usepackage{amssymb}

\newcommand{\rmo}{$R$MnO$_3$}
\newcommand{\nmo}{NdMnO$_3$}
\newcommand{\pmo}{PrMnO$_3$}
\newcommand{\tmo}{TbMnO$_3$}

\newcommand{\gfo}{GdFeO$_3$}

\newcommand{\lmo}{LaMnO$_3$}
\newcommand{\ngo}{NdGaO$_3$}
\newcommand{\kp}{\kappa}
\newcommand{\al}{\alpha}
\newcommand{\neel}{N\'{e}el}
\newcommand{\TN}{$T_{\rm N}$}
\newcommand{\Rg}{\textit{R}\,=\,}
\newcommand{\rfs}{Refs.~}
\newcommand{\rf}{Ref.~}

\newcommand{\Fig}{Figure~}
\newcommand{\equ}{equation~}
\newcommand{\T}{\,\mbox{T}}
\newcommand{\K}{\,\mbox{K}}

\newcommand{\alc}{\alpha_c}
\newcommand{\ala}{\alpha_a}
\newcommand{\alb}{\alpha_b}

\newcommand{\kpa}{\kappa_a}
\newcommand{\kpb}{\kappa_b}
\newcommand{\Hpa}{\mbox{$H\parallel a$}}
\newcommand{\Hpb}{\mbox{$H\parallel b$}}
\newcommand{\Hpc}{\mbox{$H\parallel c$}}

\newcommand{\JNF}{$J_{\parallel}^{\rm FM}$}
\newcommand{\JNA}{$J_{\perp}^{\rm AFM}$}
\newcommand{\JNN}{$J_{\parallel}^{\rm AFM}$}
\newcommand{\TNM}{$T_{\rm N}^{\rm Mn}$}
\newcommand{\D}{J_{\rm DM}}
\newcommand{\TFE}{$T_{\rm FE}$}
\newcommand{\etal}{{\it et~al.}}
\newcommand{\afm}{antiferromagnetic}
\newcommand{\fm}{ferromagnetic}

\newcommand{\bltrdw}{$\blacktriangledown$}

\newcommand{\opci}{{\large$\circ$}}
\newcommand{\optrup}{{\small$\bigtriangleup$}}

\newcommand{\NA}{N_{\rm A}}
\newcommand{\muB}{\mu_{\rm B}}
\newcommand{\kB}{k_{\rm B}}
\newcommand{\MWF}{M_{\rm WF}}

\begin{document}

\title{Anomalous thermal expansion and strong damping of the thermal
conductivity \\ of  NdMnO$_3$ and TbMnO$_3$ due to $4f$
crystal-field excitations}

\author{K.~Berggold}
\author{J.~Baier}
\author{D.~Meier}
\author{J.A.~Mydosh}
\author{T.~Lorenz}


\affiliation{$ $II. Physikalisches Institut, Universit\"{a}t zu K\"{o}ln,
Z\"{u}lpicher Str. 77, 50937 K\"{o}ln, Germany}
\author{J.~Hemberger}
\affiliation{$ $Experimentalphysik V, Institut f\"{u}r Physik, University of Augsburg, 86135 Augsburg, Germany}
\author{A.~Balbashov}
\affiliation{Moscow Power Engineering Institute, 105835 Moscow, Russia}
\author{N.~Aliouane and D.N.~Argyriou}
\affiliation{Hahn-Meitner-Institut Berlin, Glienicker Str. 100,
D-14109 Berlin, Germany}

\date{\today}
\begin{abstract}
We present measurements of the thermal conductivity $\kp$ and the
thermal expansion $\al$ of \nmo\ and \tmo. In both compounds a
splitting of the $4f$ multiplet of the $R^{3+}$ ion causes
Schottky contributions to $\al$. In \tmo\ this contribution
arises from a crystal-field splitting, while in \nmo\ it is due
to the Nd--Mn exchange coupling. Another consequence of this
coupling is a strongly enhanced canting of the Mn moments. The
thermal conductivity is greatly suppressed in both compounds. The
main scattering process at low temperatures is resonant
scattering of phonons between different energy levels of the $4f$
multiplets, whereas the complex $3d$ magnetism of the Mn ions is
of minor importance.
\end{abstract}
 \maketitle
\section{Introduction}
The search for magnetoelectric materials with the possibility to
influence magnetic (electric) ordering by an electric (magnetic)
field has greatly increased the interest in so-called multiferroic
materials, in which magnetic and ferroelectric ordering phenomena
coexist\cite{kimura03a}. The orthorhombic rare-earth manganites
\rmo\ are particularly important here, since for \Rg\,Gd, Tb, and
Dy a ferroelectric phase develops within a magnetically ordered
phase. These compounds show complex magnetic structures driven by
frustration effects, and there is evidence that the ferroelectric
order is related to a cycloidal magnetic ordering. Since the
different magnetic and electric phase transitions strongly couple
to lattice degrees of freedom~\cite{baier06a,baier06b,meier07a},
one may expect a strong influence of these ordering phenomena and
the related low-lying excitations on the phonon thermal
conductivity. In fact, recent zero-field thermal conductivity
measurements of various \rmo\ compounds by
Zhou~\etal~\cite{zhou06a} seem to support such a view. The thermal
conductivity is found to be drastically suppressed for \Rg\,Tb and
Dy, while it shows a rather conventional behavior for the other
\rmo\ compounds. Thus, the authors of Ref.~\onlinecite{zhou06a}
interpreted the suppressed thermal conductivity of \Rg\,Tb and Dy
as a consequence of the complex ordering phenomena in these
compounds. Based on the above conjecture we have studied the
magnetic-field influence on the thermal conductivity of
multiferroic \tmo\ and of \nmo\ which shows a more conventional
antiferromagnetic order. A detailed analysis of the thermal
conductivity in combination with results from thermal expansion
measurements of \nmo\ and \tmo\ reveals that for both compounds
the suppression of the thermal conductivity is largely determined
by resonant scattering of phonons by different energy levels of
the $4f$ orbitals. Therefore, our new results are in strong
contrast to the interpretation proposed in
\rf~\onlinecite{zhou06a}.

The starting compound of the \rmo\ series, \lmo, crystallizes in
an orthorhombic crystal structure (Pbnm) with a \gfo\ type
distortion. If La is replaced by smaller rare-earth ions the \gfo\
type distortion increases, which causes a decreasing Mn-O-Mn bond
angle. In \lmo,  a Jahn-Teller ordered
state\cite{rodriguezcarvajal98a} is realized below $T_{\rm
JT}\approx750$~K, and an A-type \afm\ (AFM) ordering of the Mn
moments develops below \TNM$\approx140$\K. This type of ordering
is characterized by a ferromagnetic alignment of the magnetic
moments within the $ab$ planes, and an antiferromagnetic one along
the $c$ axis\cite{wollankoehler55a,goodenough61a,ishihara97a}. A
Dzyaloshinski-Moriya (DM) type interaction $\D$ causes a canting
of the spins towards the $c$ direction resulting in a weak
ferromagnetic moment ($\MWF$). This A-type AFM ordering remains
for \Rg\,Pr,\ldots,Eu, but the \neel\ temperature is successively
suppressed. There are three main exchange couplings between the Mn
moments: the \fm\ nearest-neighbor (NN) coupling (\JNF) within the
$ab$ plane, the next-nearest neighbor (NNN) \afm\ exchange
interaction (\JNN) within the $ab$ plane, and the \afm\ NN
interaction \JNA\ along the $c$ direction. A larger distortion,
i.e.\ a decreasing Mn-O-Mn bond angle, suppresses \JNF, whereas
\JNN\ hardly changes\cite{kajimoto05a,moussa96a,hirota96a}. The
increasing frustration between \JNF\ and \JNN\ destabilizes the
A-type AFM ordering and finally leads to complex ordering
phenomena for \Rg\,Gd, Tb, and
Dy\cite{kimura03a,arima05a,aliouane06a,kimura05a,kenzelmann05a,kimura03b,quezel77a,zukrowski03a,goto05a,baier06a}.
For \Rg\,Dy$\ldots$Lu,  \rmo\ crystallizes either in a \gfo-type
or in a hexagonal structure depending on the growth technique,
while for \Rg\,Er, \ldots, Lu usually the hexagonal structure is
realized\cite{hexagonal,kappahexagonal}.

The presentation of our results in the subsequent sections is
organized as follows. After a description of the experimental
setup we first concentrate on the zero-field data obtained on \nmo
. Then we discuss the influence of a magnetic field applied either
along the $c$ axis or within the $ab$ plane of \nmo. In the last
subsection our results obtained on \tmo\ will be analyzed.

\section{Results and Discussion}

\subsection{Experimental}

The NdMnO$_3$ single crystal used in this study is a cuboid of
dimensions $1.65\times 1.85\times 1.2$\,mm$^3$ along the $a$, $b$,
and $c$ direction, respectively. It was cut from a larger crystal
grown by floating-zone melting\cite{kadomtseva05a}. Magnetization
and specific heat data of the same crystal are reported in
Ref.~\onlinecite{hemberger04a}. The measurements on TbMnO$_3$ have
been performed on different small single crystals which were cut
from a larger crystal grown by floating-zone melting in an image
furnace~\cite{aliouane06a}. The thermal conductivity was measured
by a standard steady-state technique using a differential
Chromel-Au+0.07$\%$Fe-thermocouple\cite{berggold06a}. For \nmo ,
we studied $\kpb $ with a heat current along the $b$ axis in
magnetic fields applied either along $a$, $b$, or $c$. All these
measurements were performed with one set of heat contacts using
either a $140$~kOe longitudinal-field or a $80$~kOe
transverse-field cryostat. For \tmo , we present measurements of
$\kpa $, i.e.\ with a heat current along the $a$ axis. Here, we
used different configurations in order to allow measurements up to
140~kOe for all three field directions. Unfortunately, the \tmo\
crystal cracked when we increased the field above 110~kOe for
$\Hpc$. This problem also occurred on another \tmo\ crystal during
our thermal-expansion measurements for the same field direction
and we suspect that the crystals break because of strong internal
torque effects\cite{meier07a}. The longitudinal thermal expansion
coefficients $\alpha_i$ have been measured along all three crystal
axes $i=a$, $b$, and $c$ using different home-built
high-resolution capacitance
dilatometers\cite{pott83a,lorenz97a,HeyerDip05}. For the
field-dependent measurements, we concentrate on $\alpha_b$ of
\nmo\ with $\Hpc$ and $\Hpb$ up to maximum fields of 140~kOe.
Measurements with $\Hpa$ were not possible due to large torque
effects. For \tmo , we only present zero-field data of $\alpha_i
$, since the field influence is discussed in detail in
\rf~\onlinecite{meier07a}.

\subsection{\nmo\ in Zero Magnetic Field}
\label{zerofield}

\begin{figure}
\begin{center}
\includegraphics[width=0.9\columnwidth,clip]{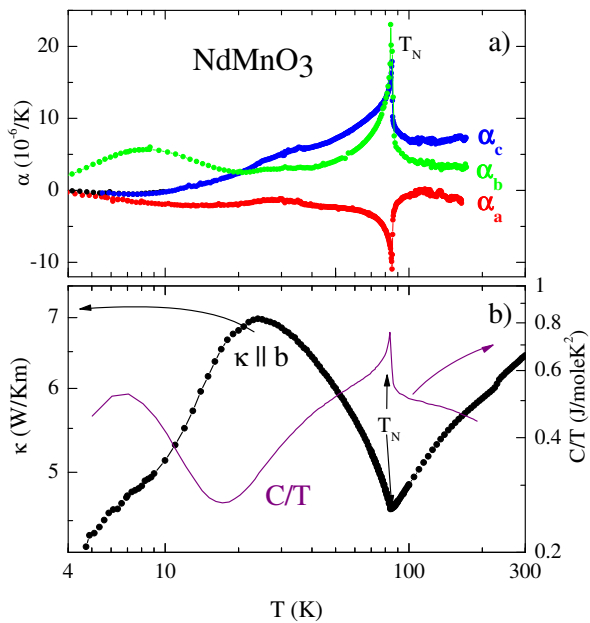} \end{center}
\caption{(Color online) a) Thermal expansion
 of \nmo\ along all
three crystallographic axes. b) Thermal conductivity measured with
a heat current along the $b$ axis (left scale) and specific heat
(right scale; data from \rf\onlinecite{hemberger04a}) of \nmo .}
\label{fig1}
\end{figure}

\Fig\ref{fig1}a shows the uniaxial thermal expansion coefficients
$\al_i$ ($i=a,b,c$) of \nmo. The \neel\ transition at \TNM$\simeq
85$\,K causes large anomalies along all three crystallographic
axes. The sign of the anomaly is positive for $\alb$ and $\alc$,
while it is negative for $\ala$. From Ehrenfest's relations it
follows that the sign of the anomaly of $\al_i$ corresponds to the
sign of the uniaxial pressure dependence of \TN . This means, for
example, \TNM\ would increase under uniaxial pressure applied
either along the $b$ or $c$ axis. The different signs of the
uniaxial pressure dependencies of \TN\ as well the suppression of
\TN\ in the \rmo\ series with increasing \mbox{Mn-O-Mn} bond angle
can be essentially traced back to the frustration between the NN
\JNF\ and the NNN \JNN\ in the $ab$ planes. For a more detailed
discussion we refer to \rfs\onlinecite{baier06b,meier07a}.

Figure~\ref{fig1}b shows the zero-field thermal conductivity
$\kpb$ (left scale) together with the specific heat (right scale;
data from \rf\onlinecite{hemberger04a}). There is a $\lambda$-like
peak in the specific heat at \TNM$\simeq 85$\,K and at the same
temperature $\kpb$ has a sharp minimum. Above \TN, $\kpb$
increases monotonically, which contradicts conventional phononic
behavior. Below \TN, $\kp$ has a maximum at $25$\K\ with a rather
low value of $7$\,W/Km. Around $8$\K\ the specific heat shows a
Schottky peak, which arises from a splitting of the $4f$
ground-state doublet of the Nd$^{3+}$ ions\cite{hemberger04a}. In
general, a two-level Schottky peak is described by
\begin{equation}\label{equ_c2niv}
C_{_{\rm sch}}=k_B\cdot\frac{\Delta^{2}}{T^{2}}\cdot \frac{\tau_{_{1}}
\tau_{_{2}} \exp(-\Delta/T)} {\left(\tau_{_{1}}+\tau_{_{2}} \exp(-\Delta/T)\right)^{2}},
\end{equation}
where $\tau_1$ and $\tau_2$ are the degeneracies of the involved
levels and $\Delta$ the energy splitting. For \nmo\ we obtain
$\Delta\simeq 21$\K\  and $\tau_1=\tau_2=1$. The $^4I_{9/2}$
ground-state multiplet of a free Nd$^{3+}$ ion is ten-fold
degenerate, but splits to five doublets in the orthorhombic
crystal field (CF). To our knowledge, no neutron scattering
investigations of the CF splitting of \nmo\ are available.
However, in the related compounds Nd$A$O$_3$ with $A=$Ni, Fe, and
Ga a splitting of the order of $200$\K\ between the ground-state
doublet and the first excited doublet has been
measured\cite{marti95a,podlesnyak93a,rosenkranz99a,przenioslo95a}.
Since a similar CF splitting is expected for \nmo , the observed
Schottky anomaly cannot arise from a thermal population of the
first excited doublet. Instead it has to be attributed to a
zero-field splitting of the Nd$^{3+}$ ground-state doublet, which
arises from the exchange interaction between the canted Mn moments
and the Nd moments\cite{hemberger04a}. A Schottky peak can also
occur in $\alpha_i$ and its magnitude and sign are given by the
uniaxial pressure dependence of the energy
gap\cite{lorenz06a,meier07a}. In \nmo , an obvious Schottky
contribution is only present for $\alb$, which will be analyzed in
detail below.

In an insulator the heat is transported by phonons. This can be
described by $\kp\propto Cv_{\rm s}\ell$, where $C$ is the
specific heat, $v_{\rm s}$ the sound velocity, and $\ell$ the mean
free path of the phonons. At low temperatures $\ell$ is determined
by boundary scattering, and $\kp$ follows the $T^3$ dependence of
the specific heat. At intermediate temperatures $\kp$ traverses a
maximum with a height strongly determined by scattering of phonons
by defects. At high temperatures $C$ approaches a constant, and
$\kp$ roughly follows a $1/T$ dependence due to Umklapp
scattering.

\begin{figure}
\begin{center}
\includegraphics[width=0.9\columnwidth,clip]{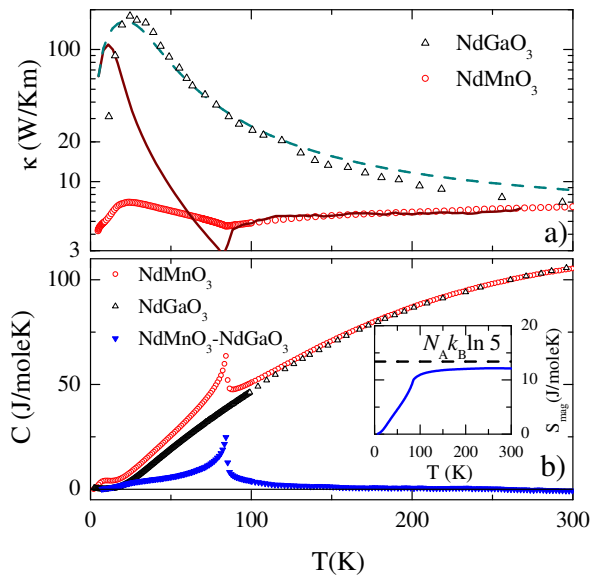} \end{center}
\caption{(Color online) a)~Zero-field thermal conductivity of
\nmo\ (\opci) and \ngo\ (\optrup) on a logarithmic $\kp$ scale.
The lines are calculated within the Debye model using the
parameters $ \Theta_D=600$~K, $v_s=6000$~m/s,
$P=3.9\cdot10^{-43}$\,s$^3$, $U=6\cdot10^{-31}$\,s/K, $u=5.6$, and
$l_{\rm min}=9$~\AA\ (see text). For the solid and dashed lines
magnetic scattering has been included
($\epsilon=4.5\cdot10^{-41}$\,m$^3$s$^3$/KJ) and excluded
($\epsilon=0$), respectively. b)~Specific heat of \nmo\ (\opci,
\rf\onlinecite{hemberger04a}) and of \ngo\ (\optrup,
\rf\onlinecite{schnelle01a}) as a reference compound. The magnetic
contribution $C_m(T)$ (\bltrdw) due to Mn spin excitations is
estimated by the difference of both curves. Inset: Magnetic
entropy $S_m=\int C_m(T)/T\,dT $ (solid line) and expected
$S_m=\NA \kB \ln(2S+1)\simeq 13.4$~J/molK (dashed) for the $S=2$
spins of Mn$^{3+}$. \label{fig2}}
\end{figure}

\Fig\ref{fig2}a compares the zero-field thermal conductivity of
\nmo\ and \ngo\cite{schnelle01a}. In \ngo, the expected
temperature dependence of conventional phononic thermal
conductivity is observed, and in the whole temperature range
$\kp$ is significantly larger than $\kp $ of \nmo . This
difference shows that additional scattering mechanisms are acting
in \nmo. For a quantitative analysis we use an extended Debye
model\cite{bermann76a}, which yields
\begin{equation}
\kappa(T) = \frac{k_{B}^4 T^3}{2 {\pi}^2 \hbar^3
  v_{s}}
\int\limits_{0}^{\Theta_{D}/T}
 \tau\left(x,T\right) \frac{x^4 e^x}{\left( e^x - 1 \right)^2}dx.
\label{kphonon}
\end{equation}
Here, $\Theta_{D}$ is the Debye temperature, $v_{s}$ the sound
velocity, $\omega$ the phonon frequency, $x = \hbar\omega/k_{B}T$,
and $\tau\left(x,T\right)$ the phonon relaxation time. The
scattering rates of different scattering mechanisms, which are
independent of each other, sum up to a total scattering rate
\begin{eqnarray}
{\tau}^{-1}(x,T) =  \frac{v_{s}}{L} + D \omega^2 + P \omega^4 + U T
\omega^3 \exp \left(\frac{\Theta_D}{uT} \right).
\label{rates}
\end{eqnarray}
The four terms on the right-hand side refer to the scattering
rates on boundaries, on planar defects, on point defects, and to
phonon-phonon Umklapp scattering, respectively. The mean free
path cannot become smaller than the lattice spacing giving a lower
limit for $\kp$. This is taken into account in \equ(\ref{rates})
by a minimum mean free path $\ell_{\rm min}$ and replacing
$\tau(x,T)$ by max\{$\tau_\Sigma(x,T),\ell_{\rm min}/v_{s}$\}
(Ref.~\onlinecite{kordonis06a}). Theoretical investigations of the
scattering of phonons by magnetic excitations yield an additional
scattering rate\cite{kawasaki63a,stern65a}
\begin{equation}
\tau_m^{-1}=\epsilon T^2C_m(T)\omega^4,
\label{equ_scatterafm}
\end{equation}
where $\epsilon$ describes the scattering strength, and $C_m$ is
the magnetic contribution to the specific heat. Note, that
fluctuations may cause a sizeable $C_m$ also above \TN.

In order to determine $C_m$ the other contributions to $C$ have
to be subtracted. In \nmo\ this background contribution $C_{\rm
bg}$ arises from acoustic and optical phonons as well as the
Schottky specific heat of the $4f$  CF excitations of Nd$^{3+}$.
Since a calculation of $C_{\rm bg}$ with the required precision
is not possible, we use \ngo\ as a reference compound. Due to the
structural similarity\cite{marti96a,mori02a}, the phonon spectrum
and the CF splitting are presumably very similar in \ngo\ and
\nmo, apart from the additional splitting of the CF doublets of
\nmo . \Fig\ref{fig2}b shows the specific heat of
\nmo\cite{hemberger04a} and \ngo\cite{schnelle01a}. At high
temperatures the curves are indeed nearly identical. We estimate
\begin{equation}
C_m(T)=C_{\text{NdMnO}_3}(T)-C_{\text{NdGaO}_3}(T)-C_{\text{sch}}^{\Delta_0}(T).
\end{equation}
The last term describes the Schottky contribution due to the
splitting of the ground-state doublet and is calculated by
\equ(\ref{equ_c2niv}) with $\Delta_0=21$\K\ and $\tau_1=\tau_2=1$.
Note that the splittings of the excited doublets do not need to be
considered, since their population sets in at higher temperature,
and and as long as the splittings are not too large they hardly
change the specific heat. In Fig.~\ref{fig2}b we display the
resulting $C_m$, which exhibits the $\lambda$ peak at \TN\ and
then slowly decays for $T>T_{\rm N}$. As a test of our analysis,
we also calculate the magnetic entropy
\begin{equation}
S_m(T)=\int_0^T\frac{C_m}{T}dT.
\end{equation}
As shown in the Inset of Fig.~\ref{fig2}b, the experimental
$S_m(T)$ approaches the expected $\NA\kB\ln(2S+1)\simeq
13.4$\,J/moleK of the Mn moments, where $\NA$ and $\kB$ denote
Avogadro's number and Boltzmann's constant, respectively.

The dashed line of Fig.~\ref{fig2}a shows the thermal
conductivity calculated for \ngo\ with the parameters given in
the figure caption. The Debye temperature and the sound velocity
have been estimated from the measured specific
heat\cite{hemberger04a} and the other parameters have been
adapted to fit the data. The calculated curve reproduces the
general behavior of $\kp$ well. In order to describe the thermal
conductivity of \nmo\, the additional magnetic scattering rate
$\tau_{m}^{-1}$ is switched on by adjusting the parameter
$\epsilon$, while keeping all the other parameters fixed. This
calculation (solid line) describes the temperature dependence of
$\kp$ above \TN\ very well. However, the calculation
overestimates the minimum at \TN , and it shows a pronounced
low-temperature maximum which is not present in the data. In
principle, the latter difference could arise entirely from
different point defect scattering in \nmo\ and \ngo . However,
our magnetic-field dependent measurements will show that this
difference arises to a large extent from an additional phonon
scattering on the CF levels.

\subsection{\nmo\ in a Magnetic Field $\Hpc$}

\begin{figure}
\begin{center}
\includegraphics[width=0.9\columnwidth,clip]{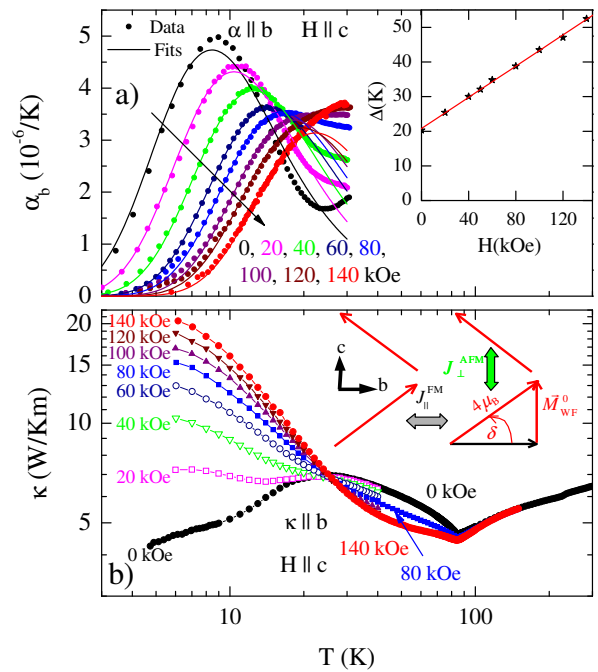} \end{center}
\caption{(Color online) a) Thermal expansion $\alb$ (symbols) of
\nmo\ for various magnetic fields $\Hpc$ together with Schottky
fits (solid lines). The arrow indicates the evolution of the
Schottky peak with increasing field. Note the logarithmic
temperature scale. Inset: Energy splitting $\Delta$ (symbols) of
the $4f$ ground-state doublet of Nd$^{3}$ as a function of
magnetic field. The line is a linear fit of $\Delta(H)$. b)
Thermal conductivity of \nmo\ for various $\Hpc$ as a function of
temperature on double-logarithmic scales. Inset: Sketch of the
canting of the Mn spins along $c$ (see text).} \label{fig3}
\end{figure}
The low-temperature Schottky contribution to the thermal
expansion of \nmo\ is most pronounced for $\alb$; see
Fig.~\ref{fig1}. For a two-level system this contribution follows
from a Gr\"{u}neisen scaling between $\al $ and the specific heat
\cite{lorenz06a}, and for two singlets it gives
\begin{equation}\label{equ_alphasch}
  \al_{\text{sch},i} =\frac{k_B}{V_{\rm uc}}\frac{\partial \ln(\Delta)}{\partial p_i}\left(\frac{\Delta}{T}\right)^2\frac{e^{-\Delta /T}}{(1 + e^{-\Delta
  /T})^2} \,\, .
\end{equation}
Here, $V_{\rm uc}$ is volume per formula unit. The magnitude and
the shape of $\al_{\text{sch},i}(T)$ are entirely determined by
the energy gap $\Delta$ and its uniaxial pressure dependence
$\frac{\partial \ln(\Delta)}{\partial p_i}$. We conclude that
$\frac{\partial \ln(\Delta)}{\partial p_i}$ is rather small for
uniaxial pressure applied along $a$ or $c$, since the Schottky
contributions of $\ala$ and $\alc$ are so small that they are
almost entirely masked by the respective phononic contributions.
In contrast, $\alb$ is clearly dominated by the Schottky
contribution $\al_{\text{sch},b}$ up to $\approx20$\K. Thus,
$\alb$ allows for a detailed analysis of the magnetic-field
dependence of the splitting of the Nd$^{3+}$ ground-state doublet.
\Fig\ref{fig3}a shows $\alb$ in magnetic fields up to $140$~kOe
applied along the $c$ direction. With increasing field, the
Schottky peak shifts monotonically to higher temperatures. The
solid lines in Fig.~\ref{fig3}a  are fits via
\equ(\ref{equ_alphasch}), which very well reproduce the
experimental data and allow to derive the energy gap as a function
of magnetic field. As shown in Fig.~\ref{fig3}b we find a linear
increase $\Delta(H)=21\,{\rm K}+2.25\cdot 10^{-4}\,{\rm K/Oe}\cdot
H$. The zero-field value nicely agrees with that obtained from the
specific heat\cite{hemberger04a}. The field dependence of $\Delta$
can be understood as follows. In zero field, $\Delta_0$ solely
arises from the Nd-Mn exchange, which is proportional to the
zero-field $\MWF ^0$. A field applied along $c$ increases $\Delta
$, on the one hand, due to the additional Zeeman splitting of the
Nd$^{3+}$ ground-state doublet. One the other hand, the canting of
the Mn moments also increases with field, yielding an additional
increase of the Nd-Mn exchange. Thus, for $\Hpc$ we obtain
\begin{equation}
\Delta(H)=\frac{\tilde{a}}{\kB}(\MWF ^0+\chi_{\rm
Mn}H)+\frac{g_{\rm Nd}\muB}{\kB} H \, . \label{equ_DcH}
\end{equation}
Here, $\tilde{a}$ is the proportionality constant between $\MWF $
and the Nd-Mn exchange, $\chi_{\rm Mn}$ the field-dependence of
$\MWF $, and $g_{\rm Nd}$ the $g$ factor of the Nd$^{3+}$
ground-state doublet.

For \lmo\ and \pmo , values of $\MWF ^0\simeq 0.1\muB$ due to the
DM interaction are reported\cite{paraskevopoulos00a,hemberger04a}.
A similar DM interaction can be expected for \nmo , but due to the
Nd-Mn exchange interaction additional energy can be gained from an
enhanced splitting of the Nd$^{3+}$ groundstate doublet by
increasing the canting of the Mn spins. In order to estimate this
effect we calculate the single-site energy of a Mn$^{3+}$ ion as a
function of the canting angle $\delta$; see Fig.~\ref{fig3}d. In a
first step, we consider the zero-field case of \pmo , where only
\JNA, $\D$, and the single-ion anisotropy $D$ of Mn$^{3+}$ have to
be considered. Based on the Hamiltonian of
Ref.~\onlinecite{moussa96a} we obtain
\begin{equation}
E^{\rm Pr}_0(\delta) =
 -4 \text{\JNA} S^2 \cos(2\delta)
 -2\D\sin\delta
 -D[S \cos\delta]^2 \label{equ_D0}
\end{equation}
where $S=2$ is the Mn spin. Using \JNA$=7$~K and $D=0.9$~K from
\rf\onlinecite{kajimoto04a} and $\delta_0=1.4^\circ $ from
$\MWF^0=4\muB\sin\delta_0=0.1\muB$\cite{hemberger04a}, we obtain
$\D =5.7$~K from the minimization condition $\partial
E/\partial\delta =0$ at $\delta_0=1.4^\circ$. In the next step we
include the additional energy gain in the Nd$^{3+}$ ground-state
doublet, which is given by $\Delta(H)/2$ from \equ(\ref{equ_DcH}),
and the potential energy of the Mn moment in a finite field \Hpc.
It is reasonable to assume that \JNA , $\D$, and $D$ do not change
from \pmo\ to \nmo\ (see also
Refs.~\onlinecite{kajimoto04a,moussa96a,hirota96a}). Thus, we keep
these parameters fixed and obtain
\begin{equation}
E^{\rm Nd}(\delta,H)=
  E^{\rm Pr}_0(\delta)
 -\frac{\Delta(H)}{2}
 - \frac{\MWF(\delta)H}{\kB} \,.
 \label{equ_etot}
\end{equation}
The determination of the remaining parameters is straightforward.
First $\partial E(\delta,H=0)/\partial\delta=0$ is solved for
$H=0$ under the additional condition that the zero-field value
$\Delta_0=21$~K is reproduced. This yields $\MWF^0=0.65\,\muB$,
$\delta_0=9.4^\circ$, and $\tilde{a}=470$~kOe. Then $\chi_{\rm
Mn}$ and $g_{\rm Nd}$ follow from a minimization of
\equ(\ref{equ_etot}) for finite fields under the condition that
the observed field dependence $\partial \Delta /
\partial H=2.25\cdot 10^{-4}\,{\rm K/Oe}$ is fulfilled.
The resulting values are $g_{\rm Nd}=2.2$ and $\chi_{\rm
Mn}=2.4\cdot10^{-6}\,\muB$/Oe. Remarkably, the weak fm moment
$\MWF^0=0.65\,\muB$ of \nmo\ is strongly enhanced compared to the
values of $\simeq 0.1\,\muB$ of \pmo\ or \lmo . This enhancement
should be clearly visible in a magnetic structure determination,
which would be a good test of our analysis.

\Fig\ref{fig3}b shows the thermal conductivity of \nmo\ up to
$H=140$~kOe applied along the $c$ direction. At $5$~K the thermal
conductivity increases almost linearly with field up to
$\kpb\simeq 20$\,W/Km. This strong field dependence weakens with
increasing temperature, and around $25$~K the field dependence
even changes sign and remains negative up $T\gtrsim T_{\rm N}$.
Our data suggest that the low-temperature behavior of $\kpb$
arises from resonant scattering of phonons between different
levels of the $4f$ multiplet of Nd$^{3+}$ which causes a
suppression of the phonon heat transport in a certain temperature
range. Such a suppression of $\kp$ by resonant scattering on $4f$
states is well-known from the literature\cite{mcclintock67a}. The
idea is that a phonon with an energy equal to the energy splitting
of two $4f$ levels is first absorbed and then reemitted. Since the
momenta of the incoming and reemitted phonons have arbitrary
directions, an additional heat resistance is caused (for more
details see e.g.\ \rf\onlinecite{hofmann01a}). The comparison of
the zero-field thermal expansion and the zero-field thermal
conductivity data gives clear evidence that resonant scattering
between the two levels of the split ground-state doublet is the
cause for the strong suppression of $\kp$ at low temperatures.
Since the splitting of the ground-state doublet increases with
increasing field the scattering probability of the low-energy
phonons systematically decreases resulting in a strong increase of
the low-temperature thermal conductivity. In the temperature range
above $\simeq 25$~K, the situation is more complex as will be
discussed in the following subsection.

\subsection{\nmo\ in Magnetic Fields $\Hpa$ and $\| b$}

\begin{figure}
\begin{center}
\includegraphics[width=0.9\columnwidth,clip]{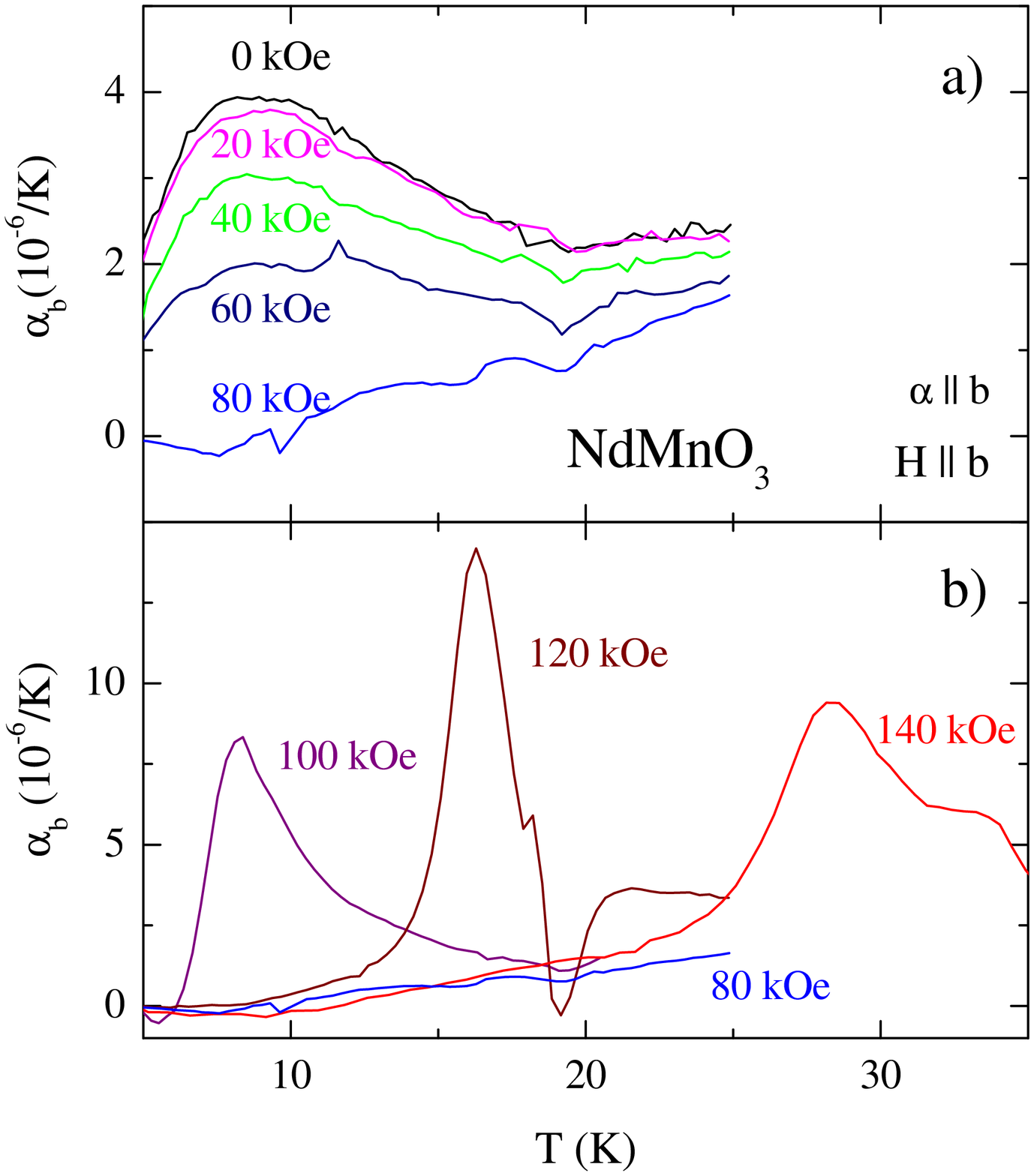} \end{center}
\caption{(Color online) Thermal expansion $\alb$ of \nmo\ with
$\Hpb$ up $80$~kOe and above 100~kOe.} \label{fig4}
\end{figure}

\Fig\ref{fig4}a shows the temperature-dependent thermal expansion
$\alb$ for magnetic fields $\Hpb$ up to $80$\,kOe. Here, the
behavior of the Schottky contribution is different to that
observed for $\Hpc$. For small fields ($H\leq20$kOe) almost no
effect occurs while for higher fields the peak height
continuously decreases until it disappears completely for
$H\simeq 80$\,kOe. The maximum of the peak weakly shifts to
higher temperature when the field is increased from 0 to
$60$\,kOe. This weak increase is a consequence of the
perpendicular orientation of the external magnetic field with
respect to the exchange field arising from $\MWF \| c$. Thus, the
total effective field is given by the vector sum of both
contributions, which for small external fields only weakly
increases. The decrease of the peak height and its disappearance
at $80$~kOe suggest that the pressure dependence of the energy gap
also decreases with field and finally vanishes. Whether this is
really the case is, however, not clear because around 100~kOe a
spin-flop transition takes place for this field
direction~\cite{hemberger04a}. Thus, different energy scales have
to be considered, which prevent a simple analysis of the pressure
dependencies via a Gr\"{u}neisen scaling~\cite{lorenz06a}. As
displayed in Fig.~\ref{fig4}b, $\alb$ for $H \geq 100~{\rm
kOe}\|b$ has another anomaly, which we attribute to the spin-flop
transition. This anomaly strongly shifts towards higher
temperatures with further increasing field, i.e.\ the phase with
$\MWF \| c$ becomes less stable towards lower temperatures.

\begin{figure}
\begin{center}
\includegraphics[width=0.9\columnwidth,clip]{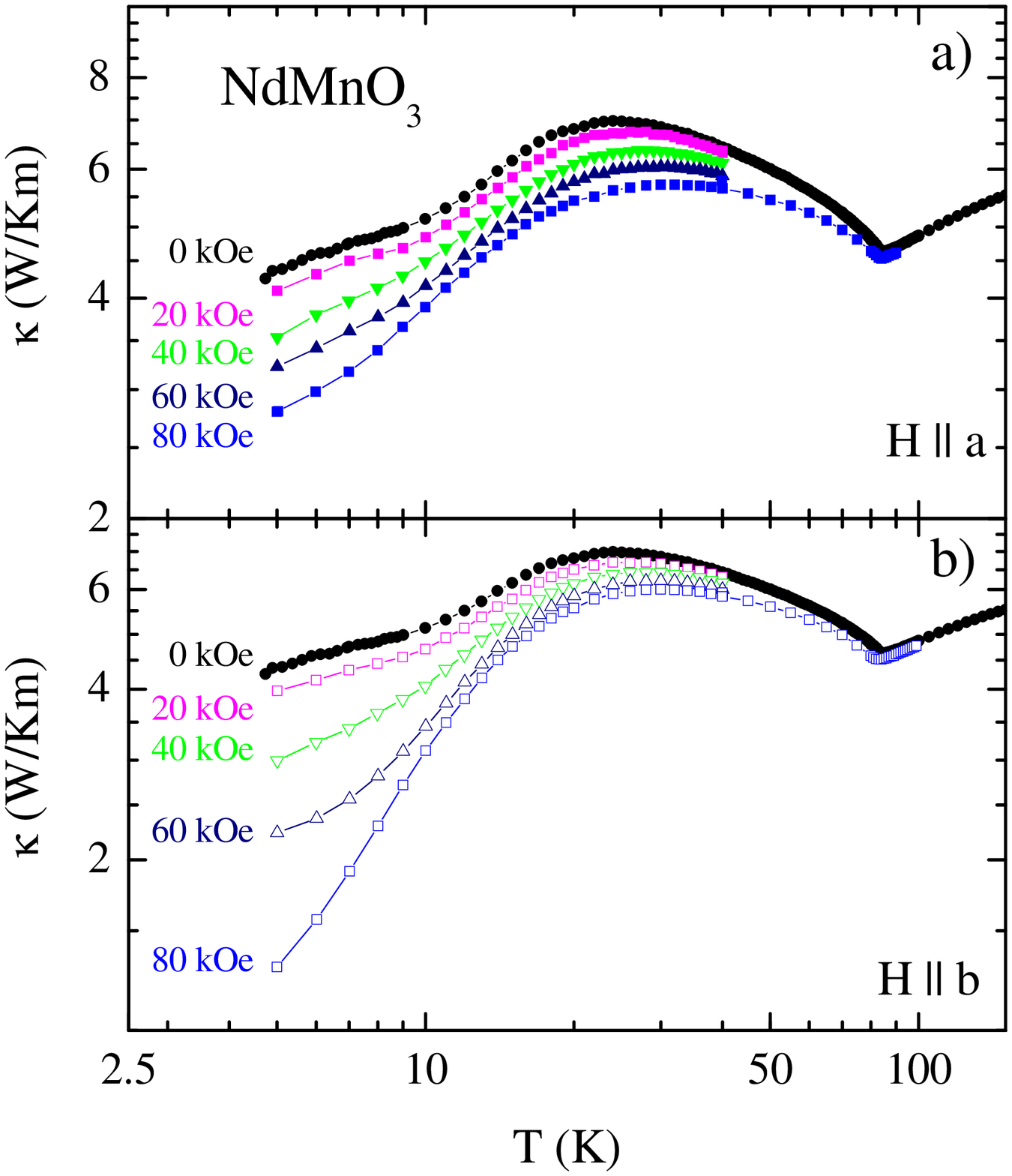} \end{center}
\caption{(Color online) Thermal conductivity $\kpb$ of \nmo\ in
magnetic fields $\Hpa$ and $\Hpb$.} \label{fig5}
\end{figure}

\Fig\ref{fig5} shows $\kp$ for $\Hpa$ and $\| b$, which is
suppressed in the entire range $T\lesssim T_{\rm N}$ for both
field directions. This behavior is in clear contrast to the strong
low-temperature increase of $\kp$ for $\Hpc$, whereas the weak
decrease of $\kp$ above about 30~K is rather similar for all three
field directions. Thus we conclude that the field dependence of
$\kpb$ is determined by two different scattering mechanisms.
Firstly, there is the resonant scattering within the split
ground-state doublet. This scattering is strongly suppressed at
low temperature for $\Hpc$ because of the increasing splitting. As
discussed above, the splitting increases much less for $\Hpa$ and
$\Hpb$, since the effective field increases only weakly for
$H\perp\MWF$. Nevertheless, a low-temperature $increase$ of $\kpb$
should occur for $\Hpa$ and $\Hpb$ if this resonant scattering was
the only field dependent scattering process. To explain the
observed decrease of $\kpb$ requires another scattering process
which increases with magnetic field for all three field
directions. This second process is only visible when the field
dependence of the first resonant one is weak, i.e., for $\Hpa$ and
$\Hpb$ in comparatively weak fields and for $\Hpc$ at $T \gtrsim
25$~K. Since the field dependence of $\kpa$ essentially vanishes
slightly above \TN\ we suspect that this second field dependent
scattering process is related to scattering of phonons by magnons,
but the presence of higher-lying CF levels of Nd$^{3+}$ could also
play a role.

\subsection{\tmo}
\tmo\ is the first compound of the \rmo\ series in which
ferroelectricity has been established over a large temperature and
magnetic-field range. The phase diagram was first explored by
polarization and magnetization data in \rf~\onlinecite{kimura05a}
and recently it has been refined by thermal expansion
measurements\cite{meier07a}. In zero field, the system transforms
from a paramagnetic to an incommensurate \afm\ phase (HTI) at
$T_{\rm N}\simeq 41$\K. At $T_{\rm FE}\simeq 27$\K, a transition
occurs into another incommensurate \afm\ phase (LTI) with a
different propagation vector. This phase is ferroelectric with a
polarization along $c$. The phase boundaries at \TN\ and $T_{\rm
FE}$ hardly depend on a magnetic field, but for $\Hpa$ and $\| b$
a transition from the LTI phase to a commensurate
antiferromagnetic phase (LTC) occurs below $T_{\rm FE}$, which is
accompanied by a polarization flop from $P\parallel c$ to
$P\parallel a$. The main difference between $\Hpa$ and $\Hpb$ is
the much larger hysteresis of the LTI-to-LTC transition for
$\Hpa$.  A magnetic field along $c$ causes a transition into a
paraelectric canted AFM phase above $\simeq 7$\T.

\begin{figure}
\begin{center}
\includegraphics[width=0.9\columnwidth,clip]{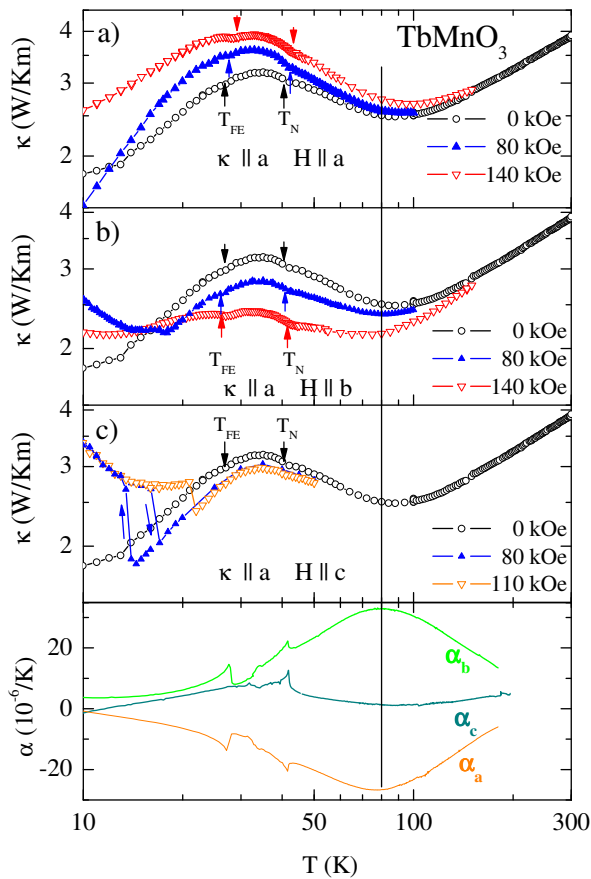} \end{center}
\caption{(Color online) a-c): Thermal conductivity $\kpa$ of
\tmo\ in magnetic fields
 along the different crystal axes. d) Thermal expansion $\al_i$ of \tmo\ with
 $i=a$, $b$, and $c$. The vertical line is a guide for the eyes.}
\label{fig6}
\end{figure}

\Fig\ref{fig6}(a-c) shows the thermal conductivity of \tmo\ along
the $a$ direction. In zero field, $\kpa$ has a broad minimum
around 80~K and a weak maximum at $\T\simeq 34$\K\ with a
relatively low absolute value $\kp\approx3$\,W/Km. \Fig\ref{fig6}d
displays the thermal expansion of \tmo\ in zero field along the
$a$, $b$, and $c$ axes. The transitions at \TN\ and \TFE\ cause
anomalies of $\al$ along all three crystallographic directions,
which are discussed in detail in \rf\onlinecite{meier07a}. In
addition, for all three directions broad Schottky contributions of
different signs are present around $80$~K, which originate from
the $4f$ states of Tb$^{3+}$. The $^7F_6$ state of the free
Tb$^{3+}$ ion splits into $13$ singlets in an orthorhombic CF. To
our knowledge, the $4f$ energy level scheme of Tb$^{3+}$ in \tmo\
is not known\cite{neutron} and we are not aware of any
investigations of the crystal-field splitting of related
Tb$A$O$_3$ compounds. Thus, a detailed analysis of this Schottky
contribution to $\al$ is not possible. As a rough estimate of the
relevant energy scale a fit of the thermal expansion data by
\equ(\ref{equ_alphasch}) yields an effective energy gap of
$\approx 190$\K. The comparison of the thermal conductivity with
the thermal expansion data shows that the extrema of $\al_i$\
occur close to the minimum of $\kp$. This correlation strongly
suggests that the minimum of $\kpa$ at $\approx90$\K\ is caused by
resonant scattering of phonons between different CF levels of
Tb$^{3+}$. Such an interpretation is further supported by the
magnetic-field dependencies of $\kpa$. For $\Hpb$ and $\Hpc$,
$\kpa$ is suppressed over a broad temperature range, whereas it is
enhanced for $\Hpa$. These large field dependencies up to high
temperatures are rather unusual, and clearly not related to the
low-temperature ordering phenomena below $41$\K. The influence of
the various transitions below 41~K on $\kpa$ is weak. We only
observe small dips at the transition temperatures \TN\ and \TFE\
as indicated by the arrows in Fig.~\ref{fig6}. The shape of the
thermal conductivity curves remains essentially unchanged. We
conclude that these ordering transitions play little role for
$\kpa$, since the low $\kpa$ is completely dominated by the
scattering of phonons by the $4f$ CF levels of Tb$^{3+}$.

Below \TFE\ additional magnetic-field dependencies are observed.
One can distinguish the measurements with $\Hpa$ or $b$, where the
LTI-to-LTC transition occurs, from those with $\Hpc$, where the
system turns into the paraelectric phase for sufficiently large
fields. For the latter case, a sharp increase of $\kpa$ is
observed when the paraelectric phase is reached, see
Fig.~\ref{fig6}c. The pronounced hysteresis of the $80$~kOe curve
reflects the first-order nature of this transition. Apart from the
different transition temperatures, the curves for $80$~kOe and
$110$~kOe are almost identical. For $H=80\text{\,kOe}\parallel b$,
$\kpa$ increases at the LTI-to-LTC transition ($\approx18$~K; see
Fig.~\ref{fig6}b). Although this field dependence is of opposite
sign compared to the high-temperature field dependence for $\Hpb$,
it is difficult to separate both effects, because the scattering
by the CF levels of Tb$^{3+}$ probably causes a magnetic-field
dependence of $\kp$ down to the lowest temperature. Since the
LTI-to-LTC transition is accompanied by a polarization flop from
$P\parallel a$ to $P\parallel c$, one may suspect that the
increase of $\kpa$ is related to the formation of ferroelectric
domains. We have, however, ruled out this possibility by
measurements of the electrical polarization as well as of the
thermal conductivity under application of large electrical fields
(not shown). Although the domain formation could be clearly seen
in the polarization measurements, no electric-field influence was
detectable in $\kp$. Thus, another explanation for the suppressed
thermal conductivity in the LTI phase is needed. Probably, it is
the incommensurability itself, which causes an additional thermal
resistance, because the crystal symmetry is lowered in the LTI
phase. Accordingly, one should expect a similar behavior for
$\Hpa$, but the broadening of the LTI-to-LTC transition due to its
large hysteresis\cite{meier07a} and the superimposed
high-temperature field dependence prevent a separation of the
different effects at low temperatures. The incommensurability is
also consistent with the jump of $\kp$ for $\Hpc$, since the
paraelectric phase is commensurate and therefore of higher
symmetry.

\section{Conclusions}
We have studied the thermal expansion and thermal conductivity of
\nmo\ and \tmo\ under application of large magnetic fields. The
thermal conductivity of \nmo\ is a very unusual. The N\'{e}el
transition at \TN$\simeq 85$\K\ leads to a strong suppression of
the phonon thermal conductivity over a large temperature range.
Including a magnetic scattering rate proportional to the magnetic
specific heat allows us to describe the thermal conductivity from
\TN\ to room temperature. At low temperatures the thermal
conductivity is further suppressed by another scattering
mechanism. The $4f$ ground-state doublet of Nd$^{3+}$ is split
($\Delta_0\approx21$\K ) by an exchange interaction with the
canted Mn moments. Our analysis suggests a significant enhancement
of the Mn canting angle in \nmo\ compared to that in \pmo\ as a
consequence of this Nd-Mn interaction. The splitting of the
ground-state doublet thereby allows for resonant scattering of
phonons which causes the additional suppression of $\kp$ in zero
field. The analysis of the thermal expansion in magnetic fields up
to $140$~kOe reveals that $\Delta(H)$ strongly increases in
magnetic fields $\Hpc$. This increase of $\Delta(H)$ shifts the
effectiveness of the resonant scattering processes towards higher
temperature and causes a drastic increase of $\kp$ at low
temperatures. For $\Hpc$, with increasing temperature a gradual
change occurs leading to a suppression of $\kp$. A similar
suppression is present for $\Hpa$ and $\Hpb$ in the entire
low-temperature range. This requires the presence of a second
field-dependent scattering mechanism, which may be related to
scattering of phonons either by magnons or by higher-lying CF
levels.

\tmo\ also exhibits a strongly suppressed thermal conductivity
over the entire temperature range. The clear correlation of the
temperature dependencies of $\kp$ and of the uniaxial thermal
expansion coefficients $\al$ enables us to conclude that the
dominant mechanism suppressing $\kp$ is resonant scattering of
phonons by the $4f$ CF levels of Tb$^{3+}$. The interpretation of
\rf\onlinecite{zhou06a} that the low absolute values of the
thermal conductivity of \tmo\ should be caused by the complex
magnetic and electric ordering phenomena is ruled out by our data.
In contrast, the complex transitions of \tmo\ only cause very weak
anomalies in $\kp$ at \TN\ and \TFE . A somewhat larger influence
is present at the transitions induced by finite magnetic fields.
The LTI-to-LTC transition for $\Hpa,b$ as well as the transition
to the paraelectric phase for $\Hpc$ cause an increase of the
thermal conductivity. Probably, this increase of $\kp$ arises from
the incommensurability of the LTI phase, which is transformed to a
commensurate phase of higher symmetry for all three field
directions. We also found that the ferroelectric domain structure
has no measurable influence on the heat transport in \tmo .

\begin{acknowledgments}
We acknowledge useful discussions with P.~Hansmann, M.~Haverkort,
D.~Khomskii, D.~Senff, and A.~Sologubenko. This work was supported
by the Deutsche Forschungsgemeinschaft through SFB 608.
\end{acknowledgments}


\begin{thebibliography}{10}
\bibitem{hexagonal}
The manganites with \Rg\,Er, \ldots, Lu have been prepared in
perovskite
  structure by high-pressure synthesis\cite{zhou06b}. Here, an E-type
  antiferromagnetic ordering occurs below $T_{\rm N}\approx 40$\K .
\bibitem{kappahexagonal}
For the thermal conductivity in the hexagonal phase other
mechanisms are
  dominant, e.g.\ rattler atoms\cite{zhou06a,sharma04a}.
\bibitem{neutron}
In \rf\onlinecite{kajimoto04a} a peak at $4$\,meV observed by
neutron
  scattering was attributed to a CF excitation, but these measurements were
  restricted to $E<10$\,meV.
\end{thebibliography}

\end{document}